\documentstyle[12pt]{article}

\def\overlay#1#2{\ifmmode%
\setbox0=\hbox{$#1$}%
\setbox1=\hbox to\wd0{\hss$#2$\hss}\else%
\setbox0=\hbox{#1}%
\setbox1=\hbox to\wd0{\hss#2\hss}\fi%
 #1\hskip-\wd0\box1 }

\begin{document}
\hfill\vbox{\hbox{NUHEP-TH-94-7}  \hbox{UCD-94-4}\hbox{hep-ph/9405261}
\hbox{April 1994}}\par
\thispagestyle{empty}

\vspace{0.5in}

\begin{center}
{\Large \bf Spin Alignment in the Production of Vector Mesons
with Charm and/or Beauty via Heavy Quark Fragmentation}

\vspace{0.3in}

Kingman Cheung\footnote{Electronic mail address:
{\tt cheung@nuhep.phys.nwu.edu}} \\

{\it Dept. of Physics \& Astronomy, Northwestern University, Evanston,
IL 60208} \\

\vspace{0.3in}

Tzu Chiang Yuan \footnote{Electronic mail address:
{\tt yuantc@ucdhep.ucdavis.edu}} \\

{\it Davis Institute for High Energy Physics \\
Dept. of Physics, University of California, Davis, CA 95616 }
\end{center}

\newpage
\begin{abstract}
We calculate the process-independent fragmentation functions for a
$\bar b$ antiquark to fragment into longitudinally and transversely polarized
$B_c^*$ ($^3S_1$) mesons to leading order in the QCD strong coupling constant.
In the special case of equal quark mass we recover previous results for the
fragmentation of $c\to \psi$ and $b\to\Upsilon$.
Various spin asymmetry parameters are defined as measures of the relative
population of the longitudinally and transversely polarized vector meson
states.
In the heavy quark mass limit $m_b \to \infty$ our polarized fragmentation
functions obey heavy quark spin symmetry, we therefore apply them as a model
to describe the fragmentation of charm and bottom into heavy-light mesons
like $D^*$ and $B^*$. The spin asymmetry parameter, $\alpha(z)$, is
consistent with the existing  CLEO data for $D^*$.
The scaling behavior  of $\langle z \rangle$ is studied in detail.
We find excellent agreement between the predictions of $\langle z \rangle$ from
our fragmentation functions and  the experimental data for $D^*$ and
$B^*$ from the LEP, CLEO, and ARGUS detectors.
Finally, we also point out that the spin asymmetry depends significantly
on the transverse momentum $p_\perp$ of the vector mesons relative to the
fragmentation axis.

\end{abstract}

\newpage
\begin{center}\section{Introduction}\end{center}
\label{intro}

The physics of hadrons containing a single heavy quark
has been studied intensively in the past several years, mainly due to the
development of the powerful technique of
Heavy Quark Effective Theory (HQET) \cite{wiseguys,hqet,review}.
One crucial observation in HQET is that the heavy quark spin decouples from
the strong interaction dynamics in the limit of infinite heavy
quark mass. This happens
because the leading operator, the chromo-magnetic dipole
moment, that couples the heavy quark spin to the gluon field
is inversely proportional to the mass of the heavy quark.

Whenever  a heavy quark is produced with high degree of polarization
(for example, bottom and charm quarks are produced with 94\% and 67\%
left-handed polarization, respectively,
at the $Z^0$ pole), it might be possible to extract the spin
information of the heavy quark if the subsequent hadronization
does not lead to substantial depolarization.
The energy spectrum or the angular distributions of the lepton in the
semi-leptonic decays of the top, bottom, or charm quarks are often used
as the spin analyzer of the heavy quark spin.
Several recent works
(Refs.\cite{closeetal,kornerkramer,mannelschuler,falkpeskin,bonran})
have been devoted to the
polarization effects in the $D^*$ and $D^{**}$ mesons, $B^*$ and $B^{**}$
mesons, and $\Sigma_b$, $\Sigma_b^*$, and $\Lambda_b$
baryons at the LEP energy.
In this article, we will
investigate in detail the spin alignment of a heavy quark that undergoes
fragmentation into the spin triplet S-wave meson state.

Recently, it has been pointed out \cite{bcfrags}
that the dominant production of $(\bar b c)$ mesons at the large transverse
momentum region is due to fragmentation,
in which a high energy $\bar b$ quark is produced from
a hard process and subsequently
fragments into various $(\bar b c)$ meson states.
Furthermore, the corresponding process-independent
fragmentation functions of
$\bar b \to B_c(^1S_0)$ and $\bar b \to B_c^*(^3S_1)$
were calculated within perturbative QCD (PQCD) \cite{bcfrags,changchen}
to leading order
in both the strong coupling constant $\alpha_s$ and $v$, where $v$ is the
typical velocity of the charm quark inside the meson.
A lower bound of the inclusive branching fraction for the production of
$B_c$ at $Z^0$ has been estimated to be about
$2.3 \times 10^{-4}$ \cite{bcfrags}, including both $1S$ and $2S$ states.
It implies that only about 230 $B_c^+$ or 230 $B_c^-$ are produced from
$10^6\;Z^0$.
At the Tevatron with a luminosity of $25 \, {\rm pb}^{-1}$,
one expects \cite{thekingandi} about $2 \times 10^4$
$B_c$ mesons to be produced with $p_T(B_c) > 10$  GeV from
the direct $\bar b$ quark and induced gluon fragmentation.
The three-charged leptons from a secondary vertex observed in the decay
$B_c^+ \to J/\psi + \bar \ell'\nu_{\ell'}\,$
followed by $J/\psi \to \ell \bar \ell$,
where $\ell,\,\ell'=e,\,\mu$,
can provide a clean signature for $B_c$. The combined branching ratio
of these decays is expected to be $\sim 0.2 \%$, which implies
40 distinct events at the Tevatron. In addition, the
$B_c$ meson can be fully reconstructed via hadronic decay modes, {\it e.g.},
$B_c^+\to J/\psi + \pi^+$, with $J/\psi \to \ell\bar \ell$.
Thus, unless LEP
can increase its luminosity by an order of magnitude or so in the
near future, the best place to look for the $B_c$ meson will be at
the Tevatron.

In Ref.~\cite{falketal}, the asymmetry due to the relative
probabilities for the production of
the transversely versus longitudinally polarized $J/\psi$ states
by $c$ and $\bar c$ quark fragmentation
was discussed.  As a result of this asymmetry, an anisotropic angular
distribution of the leptons in the decay $J/\psi \to \ell \bar \ell$
was found to be of order 5\%, which might not be large enough to be observed
due to the presence of another important source of highly polarized
$J/\psi$  from $B$-meson decays. Since the heavy quark fragmentation
probabilities, first calculated in \cite{clavelli},
are known to have big cancellation in the equal
mass quarkonium case, one expects to have larger asymmetry
in the unequal mass case like the $B^*_c$ meson system.
Such asymmetry can be
measured in principle from the anisotropy of the photon angular distribution
in the decay $B_c^* \to B_c + \gamma$.
To predict the degree of anisotropy one needs to know
the polarized fragmentation functions for the $\bar b$ quark splitting into
various helicity states of the $B_c^*$ meson.

In this work, we will study the spin alignment in the $B_c^*$ production via
the fragmentation of the $\bar b$ quark.
First, we will calculate the fragmentation functions for
$\bar b \to$ polarized $B_c^*$ states, as an extension
to the previous calculation of $\bar b \to$ unpolarized $B_c^*$
\cite{bcfrags}, or as an unequal mass extension to the calculation
of $c$ (or $\bar c$) $\to$ polarized $J/\psi$ \cite{falketal}.
This extension is also useful beyond the $(\bar b c)$ system as
the final results of the fragmentation functions can be applied
phenomenologically to the case of heavy-light mesons such as
$D^*$ and $B^*$. This allows us to get a better insight
to the spin asymmetry in the $D^*$ and $B^*$ mesons,
where perturbative methods are  not applicable.

Spin alignment of the $B_c^*, J/\psi, \Upsilon, D^*$, and $B^*$
systems will be studied.
Two spin asymmetry parameters frequently quoted in the
literature are
\begin{equation}
\label{xi}
\xi=\frac{T}{L+T} \qquad \qquad {\rm and} \qquad
\qquad \alpha=\frac{2L-T}{T}\,,
\end{equation}
where $T(L)$ denotes the production probability of
the transverse (longitudinal) state of the excited S-wave meson
($B_c^*, J/\psi, \Upsilon, D^*$, or $B^*$).
Here we introduce  another parameter
\begin{equation}
\label{W}
{\cal W} = \frac{T}{T + 2 L} \; .
\end{equation}
Two useful relations of these spin asymmetry parameters are the following
\begin{equation}
{\cal W} = \frac{\xi}{2-\xi} \qquad , \qquad
\alpha = \frac{2-3\xi}{\xi} \; .
\end{equation}
Note that we did not specify the production mechanism in the definitions
of these spin asymmetry parameters.

To measure these spin asymmetry parameters, one can study
the two body decay of the excited meson, {\it e.g.} $B^*\to B\gamma$,
$D^*\to D\pi$, and $B_c^* \to B_c \gamma$ etc.
The angular distribution of the emitted photon (or pion)
depends on the helicity of the parent meson. For
definiteness we will consider the $B_c^*$ meson in the following.
Suppose the $B_c^*$ meson is polarized in such a way that a fraction
$\xi$ is transverse while a fraction $(1-\xi)$ is longitudinal.
The transversely polarized component of the $B_c^*$
gives rise to an angular distribution of $(1+\cos^2\theta)/4$
weighted by the relative probability $\xi$,
while the longitudinal polarized component has an angular
distribution of $(1-\cos^2\theta)/2$
weighted by the relative probability $(1-\xi)$, where
$\theta$ is the angle between the outgoing photon 3-momentum and
the polarization axis in the $B_c^*$ rest frame. We define the
polarization axis to be the direction of the 3-momentum of the $B_c^*$
in the laboratory  frame.
Summing over all the helicity states of the $B_c^*$,
the angular distribution of the emitted photon is given by
\begin{eqnarray}
\label{angulardis}
\frac{d\Gamma}{d \cos \theta} &\sim &
1+ \left(\frac{3\xi - 2}{2- \xi } \right ) \cos^2\theta \; ,  \\
&=&1- \left(\frac{\alpha}{2+\alpha} \right ) \cos^2\theta \; ,  \\
&=&1 +(2{\cal W}-1)\cos^2\theta \; . \label{junk}
\end{eqnarray}
If the relative probability $T/L$ equals $2$ as suggested by
the naive spin counting rule, $\xi$ equals $2/3$ and the resulting decay
angular distribution will be flat.  In other words, there is no
spin alignment of the $B_c^*$ meson. We will show that the above
isotropic scenario is only true in the heavy quark mass
limit and this limit is broken by the finite charm and bottom quark masses.

This paper is organized as follows.
In Section II, we will derive the polarized fragmentation functions
$D^{L,T}_{\bar b \to B_c^*}(z,s)$ that depend on both the usual fragmentation
variable $z$ and the variable $s$ which measures
the virtuality of the fragmenting $\bar b$ quark.
Covariant expressions for the transverse
and longitudinal polarization sums for massive spin 1 objects
will be derived so that covariant calculation can be
performed.
In Section III, we will discuss the heavy quark mass limit of
$m_b \gg m_c \gg \Lambda_{QCD}$. We show that our polarized fragmentation
functions satisfy heavy quark spin symmetry in this limit,
and that heavy quark symmetry breaking arises from the next-to-leading
term in the heavy quark mass expansion.
In Section IV, we will study the spin asymmetry parameters
$\xi,$ $\alpha$, and $\cal W$ in detail. We will point out that the
anisotropic angular distribution in the two body decay
of the excited meson ($D^*$, $B^*,$ or $B_c^*$) first
arises at the next-to-leading term in the heavy quark mass expansion.
We will also introduce the
$z$-dependence in the spin asymmetry parameters
$\xi(z),$ $\alpha(z)$, and ${\cal W}(z)$, and study their variations with $z$.
In Section V, we will discuss the scaling behaviors of the
mean longitudinal momentum fraction
$\langle z \rangle$ for the $B_c^*$ meson, for quarkonia,
and for the heavy-light excited mesons.
In Section VI, we will present the results of the fragmentation functions
that depend on both the variable $z$ and the transverse momentum
$p_\perp$ of the $B_c^*$ meson with respect to the fragmentation axis,
which is defined as the direction of the 3-momentum of the fragmenting $\bar
b$ quark in the laboratory frame.
Spin asymmetry parameters depending  on $p_\perp$ are also introduced
and studied in detail.
We conclude in Section VII.

\begin{center}\section{Fragmentation Functions for Polarized $B_c^*$ Meson}
\end{center}

The derivation of the polarized fragmentation functions of
$\bar b \to B_c^*$ follows closely to the unpolarized case \cite{bcfrags},
but requires  separate contributions from
the longitudinal and transverse components of the $B_c^*$ meson.
For the unpolarized case  all the helicity states are summed
by the following formula
\begin{equation}
\label{sum-un}
\sum_{\lambda} \epsilon^{*\mu}_\lambda(p)\; \epsilon^{\nu}_\lambda(p)
 = -g^{\mu\nu} +\frac{p^\mu p^\nu}{M^2}\,,
\end{equation}
where $p$ and $M$ are the momentum and mass of the meson, respectively.
Since we are working entirely within the nonrelativistic approximation
for the heavy quark bound state, we will take $M=m_b+m_c$.
The two transverse helicity states are usually summed
by using the following equation
\begin{equation}
\sum_{T} \epsilon^{*i}_T(p)\; \epsilon^j_T(p)
 = \delta^{ij}  - \frac{p^i p^j}{|\vec p|^2} \,.
\end{equation}
This noncovariant expression for summing the transverse polarizations
often makes  the manipulation unnecessarily cumbersome.
Here we present covariant formulas for summing the longitudinal
and transverse helicity states of massive spin 1 objects.
Recall that the longitudinal polarization 4-vector $\epsilon_L(p)$
is  usually written explicitly as
\begin{equation}
\epsilon^\mu_L(p) = \left( \frac{|\vec p|}{M},\; \frac{E \vec p}{M|\vec
p|}\right )\,,
\end{equation}
where $E^2=M^2 + |\vec p|^2$.  Let us define an auxiliary 4-vector
$n^\mu=(1,\;-\vec p/|\vec p|)$ such that $n^2=0$ and $n\cdot p=E+|\vec p|$.
With the help of this auxiliary vector  one can rewrite $\epsilon^\mu_L(p)$
in the following covariant form,
\begin{equation}
\epsilon^\mu_L(p) = \frac{p^\mu}{M} - \frac{Mn^\mu}{n \cdot p}\,.
\end{equation}
We can then obtain the following covariant expressions
\begin{equation}
\label{sum-l}
\epsilon^{*\mu}_L(p) \epsilon^\nu_L(p) = \frac{p^\mu p^\nu}{M^2} -
\frac{1}{n\cdot p}( p^\mu n^\nu + p^\nu n^\mu) + \frac{M^2}{(n\cdot p)^2}
n^\mu n^\nu\,,
\end{equation}
for the longitudinal polarization sum,  and
\begin{equation}
\label{sum-t}
\sum_T \epsilon^{*\mu}_T(p) \epsilon^\nu_T(p) = - g^{\mu\nu} +
\frac{1}{n\cdot p}( p^\mu n^\nu + p^\nu n^\mu) - \frac{M^2}{(n\cdot p)^2}
n^\mu n^\nu\,,
\end{equation}
for the transverse polarization sum.
In general, these covariant formulas are not only useful
in our calculation but also in  many other applications.

The initial fragmentation functions $D^L_{\bar b\to B_c^*}(z,\mu_0)$
and $D^T_{\bar b\to B_c^*}(z,\mu_0)$ are obtained by modifying
the calculation in Ref.~\cite{bcfrags}
using Eqns.~(\ref{sum-l}) and (\ref{sum-t}) to separate the
contributions from the longitudinal and transverse components of $B_c^*$.
The starting point is the following expression \cite{bcfrags}
\begin{equation}
D_{\bar b \to B_c^*}(z)
\;=\; \int ds
\; \theta \left( s - {M^2 \over z} - {m_c^2 \over 1-z} \right)
D_{\bar b \to B_c^*}(z,s) \; ,
\label{Dfragz}
\end{equation}
with
\begin{equation}
D_{\bar b \to B_c^*}(z,s)
\;=\; {1 \over 16 \pi^2} \;
\lim_{q_0/m_b \rightarrow \infty} {|{\cal M}|^2 \over |{\cal M}_0|^2} \;.
\label{Dfragzs}
\end{equation}
In Eqn.(\ref{Dfragzs}), ${\cal M}$ is the amplitude for producing
a $B_c^*$ and a $\bar c$
quark  from an off-shell $\bar b^*$ with virtuality $s=q^2$,
where $q$ is the 4-momentum of the $\bar b$ quark (the leading contribution
is an one-gluon exchange diagram given in Fig.~\ref{fig-feynman});
and ${\cal M}_0$ is the amplitude for producing an on-shell $\bar b$
with the same 3-momentum $\vec q$.  If
the momentum of the $B_c^*$ is $p^\mu=(p_0,p_1,p_2,p_3)$, in a frame where
$q^\mu=(q_0,0,0,q_3)$, the longitudinal momentum fraction $z$ is defined by
$z=(p_0+p_3)/(q_0+q_3)$.  The  kinematical relation among the three
variables $z$, $s$, and $p_\perp = |\vec{p}_\perp|$, where
$\vec{p}_\perp=(p_1,p_2)$ is  the transverse momentum of the $B_c^*$ meson
with respect to the fragmentation axis, is
\begin{equation}
\label{perp}
s = \frac{M^2 + p_\perp^2}{z} + \frac{m_c^2 + p_\perp^2}{1-z} \, .
\end{equation}
Note that the $\theta$ function constraint in Eqn.~(\ref{Dfragz})
arises from the positivity of $p_\perp$.

The amplitude ${\cal M}$ is evaluated in the axial gauge
with an auxiliary 4-vector $n^\mu$ defined above in the covariant
polarization sums.
In this gauge, the dominant contribution arises from
the Feynman diagram depicted in Fig.~\ref{fig-feynman}.
Other diagrams are
suppressed by powers of $m_{b,c}/q_0$. In other words, factorization is
manifest in this gauge \cite{bcfrags}. The amplitude ${\cal M}$ is given by
\begin{eqnarray}
i {\cal M} &=&
\sqrt{\pi} \alpha_s C_F R(0)
\frac{\delta_{ij}}{\sqrt{N_c}}
\frac{\sqrt{M}}{m_c}
\frac{1}{(s-m_b^2)^2} \nonumber \\
&&\times \bar \Gamma \left ( 2M(\overlay{/}{q}+m_b)
\overlay{/}{\epsilon}^*_\lambda(p) + \frac{s-m_b^2}{n \cdot (q-\bar r p)}
(\overlay{/}{p}+M) \overlay{/}{\epsilon}^*_\lambda (p) \overlay{/}{n} \right )
v(p') \, ,
\label{eq16}
\end{eqnarray}
where $p$ and $p'$ are the momentum 4-vector of the outgoing $B_c^*$ and $\bar
c$ respectively, $\epsilon_\lambda(p)$ is the polarization 4-vector of the
$B_c^*$ meson with helicity $\lambda$, $R(0)$ is the radial wavefunction of the
$(\bar b c)$ bound-state at the origin, $M=m_b+m_c$ is the mass of the bound
state, $r=m_c/M$, $\bar r = 1-r$, and $\Gamma$ is a symbolic
Dirac structure representing the source to create the
energetic $\bar b$ quark in the hard subprocess.
In Eqn.~(\ref{eq16}),  $C_F = (N_c^2-1)/2N_c$, where $N_c$ is the
number of color.  $R(0)$ can be determined by a potential model
calculation \cite{eichtenquigg}
or extracted from the $B_c$ decay constant $f_{B_c}$. The latter can be
calculated on a lattice or measured in the future experiments.
The relation between $R(0)$ and $f_{B_c}$ is given by
$|R(0)|^2 = \pi M_{B_c} f_{B_c}^2 /3$.
Squaring the amplitude ${\cal M}$, summing over the  color and
spin of the $\bar c$ quark, and using Eqn.~(\ref{sum-l}) to project the
longitudinal component of the $B_c^*$ meson, we get
\begin{equation}
\sum |{\cal M}|^2 = \pi \alpha_s^2 C_F^2 |R(0)|^2
\frac{M^5}{m_c^2}\frac{1}{(s-m_b^2)^4} {\rm Tr}
(\bar \Gamma\Gamma \overlay{/}{q})
\times \Delta^L(z,s) \,,
\end{equation}
where $\Delta^L(z,s)$ is given by
\begin{eqnarray}
\Delta^L(z,s) &=& \frac{16}{z^2} \left( 1 - 2 \bar r z
+ \bar r (1-2r)z^2 \right ) \nonumber \\
&+& \frac{4(s-m_b^2)}{z(1-\bar r z)M^2} \left( -4+2(3-4r)z -(1-8r+4r^2)z^2
-\bar r (1-2r)z^3 \right ) \nonumber \\
&+& \frac{4(1-z)(1+rz)^2(s-m_b^2)^2}{(1-\bar r z)^2M^4} \; .
\end{eqnarray}
To arrive at this expression we have made the
substitutions of $p=zq$ and $p'=(1-z)q$ at the final step,
which are accurate to leading order in $m_{b,c}/q_0$.
In the fragmentation limit of $q_0/m_b \to \infty$,
the tree level amplitude ${\cal M}_0$ is simply
\begin{equation}
\sum |{\cal M}_0|^2 = N_c {\rm Tr}(\bar \Gamma \Gamma \overlay{/}{q})\,.
\end{equation}
Thus we obtain
\begin{equation}
D^L_{\bar b\to B_c^*}(z,s) = \frac{3}{2} rNM^6
\frac{\Delta^L(z,s)}{(s-m_b^2)^4}\, ,
\end{equation}
where we have defined
\begin{equation}
\label{norm}
N = \frac{\alpha_s^2(2m_c) C_F^2 |R(0)|^2}{24 N_c \pi m_c^3} \, .
\end{equation}
The scale of the strong coupling constant has been set to be $2 m_c$ --
the minimal virtuality of the exchange gluon \cite{bcfrags}.
Doing the $s$ integration,  we obtain the longitudinal
fragmentation function
\begin{eqnarray}
\label{dzl}
D^L_{\bar b\to B_c^*}(z, \mu_0) =& N &
\frac{rz(1-z)^2}{(1-\bar r z)^6}
\Bigg[ 2 + 2(2r-3)z + (16r^2 -10r +9)z^2  \nonumber \\
& - & 2 \bar r (6r^2 -5r+4)z^3 +3\bar r^2 (2r^2-2r+1)z^4 \Bigg] \,,
\end{eqnarray}
where the initial scale $\mu_0$ has been set to be $(m_b+2m_c)$
-- the minimal virtuality of the fragmenting $\bar b$ quark \cite{bcfrags}.

Similarly, one can use Eqn.~(\ref{sum-t}) to project out the transversely
polarized $B_c^*$ state and deduce
\begin{equation}
D^T_{\bar b\to B_c^*}(z,s) = \frac{3}{2} rNM^6
\frac{\Delta^T(z,s)}{(s-m_b^2)^4}\,,
\end{equation}
with
\begin{eqnarray}
\Delta^T(z,s)= & - & \frac{16}{z^2} \left( 1 - 2\bar r z + \bar r
(1+r) z^2 \right )
\nonumber \\
& + & \frac{8(s-m_b^2)}{z(1-\bar r z)M^2} \left( 2 + 2(r-2)z +(2r+1)z^2
+\bar r z^3 \right ) \nonumber \\
& + & \frac{8z^2(1-z)(s-m_b^2)^2}{(1-\bar r z)^2M^4}  \; .
\end{eqnarray}
Doing the $s$ integration, we obtain the transverse fragmentation function
\begin{eqnarray}
\label{dzt}
D^T_{\bar b\to B_c^*}(z, \mu_0) &=& 2N
\; \frac{rz(1-z)^2}{(1-\bar r z)^6}
\Bigg[ 2 + 2(2r-3)z + (10r^2 -4r +9)z^2  \nonumber \\
&& \qquad \qquad \qquad \; \; - 2 \bar r (r+4)z^3 +3\bar r^2 z^4 \Bigg] \, .
\end{eqnarray}
At this point, a few cross checks can be made.
Adding  $D^L_{\bar b\to B_c^*}(z)$
and $D^T_{\bar b\to B_c^*}(z)$, we reproduce the unpolarized
result given in Refs.~\cite{bcfrags,changchen}.
By setting $r=1/2$ in the
Eqns.~(\ref{dzl}) and (\ref{dzt}), we reproduce the results of
$c \to$ polarized $J/\psi$ given in Ref.~\cite{falketal}.
The results given in
Eqns.~(\ref{dzl}) and (\ref{dzt}) also agree with a recent calculation
of Ref.\cite{chen}.

One can easily extend these results to the case where the
initial fragmenting  $\bar b$ quark is also polarized. Let us denote
$D_{(h;\lambda)}(z) $ to be the fragmentation function for a heavy quark
$Q$ with helicity  $h = \pm 1/2$ to split into a vector meson $V^*$ with
helicity $\lambda = 0,\pm$. Thus, by definition, we have
\begin{equation}
D^L (z) = \frac{1}{2} \left(
D_{({1 \over 2};0)}(z) + D_{(-{1 \over 2};0)}(z) \right )\; ,
\end{equation}
and
\begin{equation}
D^T (z) = \frac{1}{2} \left( D_{({1 \over 2};+)}(z) + D_{({1 \over 2};-)}(z) +
          D_{(-{1 \over 2};+)}(z) + D_{(-{1 \over 2};-)}(z) \right )\; .
\end{equation}
Parity invariance implies
\begin{equation}
D_{(h;\lambda)}(z) = D_{(-h;-\lambda)}(z) \; .
\end{equation}
Therefore, we deduce the following relations
\begin{equation}
D_{({1 \over 2};0)}(z) =
D_{(-{1 \over 2};0)}(z) = D^L(z) \; ,
\end{equation}
and
\begin{equation}
D_{({1 \over 2};+)}(z) + D_{({1 \over 2};-)}(z) =
D_{(-{1 \over 2};+)}(z) + D_{(-{1 \over 2};-)}(z)
= D^T(z) \; .
\end{equation}
These relations immediately imply that the polarized heavy quark fragmentation
functions $D^L(z)$ and $D^T(z)$ are the same whether the initial heavy quark
is polarized or unpolarized.

\begin{center}\section{Heavy Quark Spin Symmetry}\end{center}

Using the technique of HQET \cite{wiseguys,hqet,review}, Jaffe and Randall
\cite{jaffe} have recently shown that the fragmentation functions
$D_{Q\to H}(z)$ for a heavy quark $Q$ to split into a hadron $H$ with one heavy
constituent quark  can be expanded as a power series in $r$,
\begin{equation}
D_{Q\to H}(z) = \frac{a(y)}{r} + b(y) + {\cal O}(r) \, ,
\end{equation}
where  $a(y)$ and $b(y)$  are functions of $y=(1-\bar r z)/rz$ and
${\cal O}(r)$ denotes all other terms higher order in $r$.
The leading term $a(y)$ is independent of the heavy quark spin and flavor;
while  the next-to-leading term $b(y)$ and all higher order terms contain
heavy quark spin-flavor symmetry breaking effects.
One can verify easily that our polarized $B_c^*$ fragmentation functions can
be expressed in this form by carefully expanding the powers of $r$ and
$(1 - \bar r z)$. The results are
\begin{eqnarray}
\label{LL}
D^L_{\bar b\to B_c^*}(z) &=& \frac{N(y-1)^2}{y^6} \Bigg[
\frac{1}{r}(8+4y+3y^2) - (8-8y+5y^2+y^3) +... \Bigg] \, , \\
\label{TT}
D^T_{\bar b\to B_c^*}(z) &=& \frac{2N(y-1)^2}{y^6} \Bigg[
\frac{1}{r}(8+4y+3y^2) - (8-8y-y^2+y^3) +... \Bigg] \,  .
\end{eqnarray}
Obviously, the leading terms of $D^L(z)$ and $D^T(z)$
in Eqns.~(\ref{LL}) and (\ref{TT}) differ only by a factor of 2 and thus
obey heavy quark spin symmetry.
The ${\cal O}(r^0)$ terms and
beyond in Eqns.~(\ref{LL}) and (\ref{TT}) are different
due to heavy quark spin-flavor symmetry breaking effects.
In fact, one can show that \cite{eff}
the leading order terms
in Eqns.~(\ref{LL}) and (\ref{TT}) can be derived by using the Feynman rules
of the leading operator in the HQET Lagrangian, while the ${\cal O}(r^0)$
pieces arise not only from the next-to-leading ($1/M$) operators
but also from the small component of the heavy quark spinor in the HQET.

In Fig.~\ref{fig-Dz}, we plot the full PQCD fragmentation functions
$D^L_{\bar b\to B_c^*}(z)$ and $D^T_{\bar b\to B_c^*}(z)$ from
Eqns.~(\ref{dzl}) and (\ref{dzt}),
and the corresponding heavy quark mass expansion from
Eqns.~(\ref{LL}) and (\ref{TT}).  We take $m_b=4.9$ GeV and $m_c=1.5$ GeV
throughout the paper.   We employ a simple form of $\alpha_s$
by evolving from its well measured experimental value at the $Z^0$-mass,
namely
\begin{equation}
\label{alphas}
\alpha_s(\mu) = \frac{\alpha_s(m_Z)}{1+  (b/2\pi)
 \alpha_s(m_Z) \log(\mu/m_Z)}
\,,
\end{equation}
where $b=(11 N_c-2 n_f)/3$, $n_f$ is the number of active flavors at the
scale $\mu$, and $\alpha_s(m_Z)=0.12$.
It is clear from Fig.~\ref{fig-Dz} that the
sum of the leading and the next-to-leading terms in the
heavy quark mass expansion is a very good approximation to the full
PQCD result, and the difference is of order
${\cal O}(r)$.  We note that the width between the peak and the endpoint of
the fragmentation functions scales as $r$.

Since our polarized fragmentation functions given in
Eqns.~(\ref{dzl}) and (\ref{dzt}) are consistent with the general
analysis of Jaffe and Randall \cite{jaffe},
we can apply them to describe the heavy-light mesons as well.
The formulas given in Eqns.~(\ref{dzl}) and (\ref{dzt}) can be
regarded as phenomenological fragmentation functions with two free
parameters $N$
and $r$, to describe the nonperturbative process of a heavy quark splitting
into a polarized heavy-light excited meson,
such as $c \to$ polarized $D^*$ and $b \to$ polarized $B^*$.
The parameter $N$ can be adjusted to describe the overall normalization and
$r$ is the mass ratio of the light constituent quark mass to the meson mass.
Although in the $D^*$ and $B^*$ systems there are probably
large nonperturbative and relativistic effects that we have not  taken
into account, our perturbative QCD fragmentation functions with the free
parameters $N$ and $r$ can at least provide some insights to these systems
while precise nonperturbative fragmentation functions for
$c\to D^*$ and $b\to B^*$ are not available yet.
Our PQCD fragmentation functions only depend on two free parameters as does
the phenomenological Peterson fragmentation function \cite{peterson}, which
is widely used in the literature to describe $c\to D^*$ and $b\to B^*$.
However, our fragmentation functions also carry spin informations.  We expect
our functions should be equally successful as phenomenological description of
fragmentation, but they are more predictive since they also contain
spin informations.  They also have an additional virtue of being rigorously
correct in some limit, namely, the limit in which the lighter quark mass is
much greater than $\Lambda_{\rm QCD}$.

Finally, we note that the 2-to-1 spin counting ratio for the transversely and
longitudinally polarized states is only true for the S-wave excited states
in the heavy quark limit. For the P-wave excited states, this statement
is incorrect \cite{falkpeskin}.

\begin{center}\section{Spin Alignment in Heavy Quark Fragmentation}
\end{center}

In this Section, we will study the spin alignment of the
excited S-wave heavy meson ($J/\psi$, $\Upsilon$,
$B_c^*$, $D^*$, or $B^*$)  produced by
heavy quark fragmentation.
To begin with, we integrate $D^{L,T}(z,\mu_0)$  over $z$ to get the total
fragmentation probabilities, or equivalently the first moments $D^{L,T}(1)$
of the corresponding fragmentation functions:
\begin{eqnarray}
\label{probl}
D^L(1) &=& \int_0^1 dz D^L (z,\mu_0) \nonumber \\
&=& N \Bigg[ {24 + 89r - 486r^2 + 354r^3 + 289r^4 \over 15\bar r^5}
\nonumber \\
&& \;\;\;\;\; \; \qquad \qquad
+ \; {r(7- 16r - 9r^2 + 30r^3 + 6r^4) \over \bar r^6} \log (r)
\Bigg] \; , \\
\label{probt}
D^T(1) &=& \int_0^1 dz D^T (z,\mu_0) \nonumber \\
&=& 2N
\Bigg[{24 + 119r +54 r^2 + 84r^3 - 11 r^4 \over 15\bar r^5}
+ \; {r(7 + 2r + 9r^2) \over \bar r^6} \log (r)
\Bigg] \; ,
\end{eqnarray}
where $r$ is defined as the mass ratio
$m_{\rm light \; quark}/m_{\rm meson}$, just the
same way as we defined for the $B_c^*$ meson.  From now on  we will
omit the subscript $\bar b \to B_c^*$ in the
fragmentation function and understood that it can refer to either
$B_c^*$, $J/\psi$,  $\Upsilon$, or the heavy-light mesons $D^*$ and $B^*$.
It is well known that the first moment of fragmentation function
has zero anomalous dimension and hence it does not evolve with the scale.
We therefore drop the $\mu$ dependence in the first moments $D^{L,T}(1)$.

Since the dominant production mechanism
of S-wave excited meson states  at the large transverse
momentum region is due to fragmentation, we can identify
the quantities $L$ and $T$, defined in Section I, to be
the first moments $D^L(1)$ and $D^T(1)$, respectively.
The various spin asymmetry parameters introduced in
Eqns.~(\ref{xi}) - (\ref{W}) can be expressed as ratios of the first moments
of the fragmentation functions as follows:
\begin{eqnarray}
\label{ximom}
\xi & = & \frac{D^T(1)}{D^T(1)+D^L(1)} \; ,\\
\label{alphamom}
\alpha & = & \frac{2 D^L(1) - D^T(1) }{D^T(1)} \; ,\\
\label{Wmom}
{\cal W} & = & \frac{D^T(1)}{D^T(1) + 2 D^L(1)}  \; .
\end{eqnarray}
Note that these spin asymmetry parameters depend only on the parameter $r$
and do not depend upon the overall constant $N$ and the evolution scale $\mu$.

Before we proceed further, it is instructive to repeat here
the usual arguments of how the heavy quark spin
information is lost during hadronization of the heavy quark into
a  S-wave heavy-light meson.  Suppose
a spin down heavy quark $Q^\downarrow$ combines with a spin up or spin down
light anti-quark $\bar q$ forming the state
$Q^\downarrow \bar q^\uparrow$ or
$Q^\downarrow \bar q^\downarrow$. Since parity is conserved in the
fragmentation process,
these two states must occur with equal probability.
While $Q^\downarrow \bar q^\downarrow$ is an eigenstate of the total spin
$S=1$; $Q^\downarrow \bar q^\uparrow$ is a mixture of the spin states $S=0$ and
$S=1$.  One can then decompose the state
$Q^\downarrow \bar q^\uparrow$ into a sum of eigenstates of the total spin
$S=0$ and $S=1$:
\begin{equation}
Q^\downarrow \bar q^\uparrow =
{1 \over \sqrt 2} \Biggl[ {Q^\downarrow \bar q^\uparrow
                          - Q^\uparrow \bar q^\downarrow \over \sqrt 2}
			\Biggr]  +
{1 \over \sqrt 2} \Biggl[ {Q^\downarrow \bar q^\uparrow
                          + Q^\uparrow \bar q^\downarrow \over \sqrt 2}
			\Biggr]  \; .
\end{equation}
These $S=0$ and $S=1$ components are identified as the
pseudoscalar $P$ and vector meson $V^*$, respectively.  In the limit of
$m_Q \to \infty$, these two states $P$ and $V^*$ are degenerate. They will
then have the same time evolution and will propagate coherently. The spin
wavefunction will remain the same as $Q^\downarrow \bar q^\uparrow$ and
the coherent superposition of the two meson states will preserve the
heavy quark spin forever.
In reality, $m_Q \neq \infty$; the pseudoscalar
$P$ and vector meson $V^*$ have a slight mass difference $\Delta M$
and hence different time evolutions. In most cases, like
the heavy-light mesons and the $(\bar b c)$ mesons,
the finite mass difference $\Delta M$ is considerably larger than
the decay rate $\Gamma_{V^*}$ of $V^* \to P + X$, where $X$ denotes a photon
or a pion.
At a time $t \sim \Delta M^{-1}$, the
$S=0$ and $S=1$ components become completely out of phase
and incoherent before any decay actually occurs.
Thus the heavy quark spin
is depolarized over a period of time characterized by the chromo-magnetic
dipole moment,
which is also responsible to the finite mass difference $\Delta M$.
At a later time $t \sim \Gamma_{V^*}^{-1}$, the vector meson $V^*$ decays
into the pseudoscalar $P$.
The spin information of the heavy quark is carried in the relative population
of the helicity $+1$ and helicity $-1$ components of the vector meson.  If
the decay proceeds through electromagnetic or strong interactions, the parity
invariance combined with rotational symmetry implies that the spin information
is lost in the angular distribution of the final state particles.
Nevertheless, the angular distribution carries important information about the
fragmentation process through the spin alignment variables, which measure the
relative population of the helicity $0$ and helicity $\pm 1$ components of the
vector meson.

In the heavy quark mass limit of
$m_b \gg m_c \gg \Lambda_{QCD}$, {\it i.e.} $r \to 0$,
the spin asymmetry parameters obtained in Eqns.(\ref{ximom})-(\ref{Wmom})
take the following simple forms
\begin{eqnarray}
\xi - {2 \over 3}&=& {5 \over 18} r + {\cal O}(r^2)  \; , \\
\alpha &=& - {5 \over 4}r + {\cal O}(r^2) \; , \\
2{\cal W} - 1 &=& {5 \over 8}r + {\cal O}(r^2) \; .
\end{eqnarray}
Note that, to order $r$,
the subleading terms ${\cal O}(r \log r)$ in Eqns.(\ref{probl}) and
(\ref{probt}) cancel in these quantities.
These limits make it clear that the angular distribution given in
Eqns.~(\ref{angulardis})--(\ref{junk}) is isotropic to leading order
in $r$. This result is in accord with
the recent general analysis of Falk and Peskin \cite{falkpeskin}
using heavy quark symmetry. As a matter of fact, in the heavy quark limit
the spin asymmetry parameter $\cal W$ coincides with the Falk-Peskin variable
$w_{1/2}$, which is the conditional probability for a
heavy quark  fragmenting into a heavy-light meson with a  spin ${1 \over 2}$
light degree of freedom to be in the  helicity states
$h_\l= + \frac{1}{2}$ or $- \frac{1}{2}$.
Light degree of freedom denotes collectively
the light quark plus all the soft gluons that
combine with the static heavy quark to form the color
singlet heavy-light system.
Parity invariance implies these two helicity configurations
$(h_\l=\pm \frac{1}{2})$ of the spin ${1 \over 2}$ light degree of freedom
must occur with equal probability \cite{falkpeskin}.
Thus $2{\cal W} \to w_{1/2} = 1$ in the heavy quark limit.
Consequently, the anisotropy of the decay product of the excited
S-wave heavy-light meson arises entirely from the spin-flavor
symmetry breaking effects of the heavy quark. Our results of the polarized
fragmentation functions and the spin asymmetry parameters obtained
above allow us to study these symmetry breaking effects
as a function of $r$.

We plot these three spin asymmetry parameters
versus $r$ in Fig.~\ref{fig-xi}(a), (b), and (c) respectively.
The anisotropy of 5.7\% \cite{falketal} in the decay distribution of
$J/\psi\to\ell\bar\ell$, where $\ell=e^-$ or $\mu^-$,
can be immediately evaluated from the values of $\xi$,
$\alpha$, or ${\cal W}$ at $r=1/2$ in
Figs.~\ref{fig-xi}.  It means that the decay distribution behaves like
$(1+0.057 \cos^2\theta)$, where $\theta$ is the angle between the outgoing
lepton and the polarization axis of the $J/\psi$.
In principle, a 5.7\%
asymmetry is measurable, but it is seriously contaminated by the $J/\psi$'s
coming from $B$ decays.
This anisotropic distribution is also true for $\Upsilon \to \ell\bar\ell$.
Surprisingly, the anisotropy in $B_c^*\to
B_c\gamma$ is only $\sim 5.8\%$, almost the
same as in the equal mass quarkonium case. But it is relatively clean
because  the production rates for the $2S, \, 2P, \, 3P,$ and $3D$
$(\bar b c)$ states
which  can contaminate the direct $\bar b \to B_c^*$ fragmentation
by their hadronic cascades or radiative  decays into $B_c^*$
are in general small. The production rates for the $2S$ $(\bar b c)$
states are about 60 \% of the corresponding $1S$ states \cite{bcfrags}.
The fragmentation probabilities for $\bar b \to$ P-wave $(\bar b c)$
states are only about 10 \% of the S-wave case \cite{bcfragp}.
While the fragmentation functions for
$\bar b \to$ D-wave $(\bar b c)$ states are not known, they are
not expected to be large.   In order to measure the asymmetry,
one still has to disentangle the issue of whether the
$B_c^*$ is coming via direct $\bar b$ fragmentation or cascade from
higher excitations.
The anisotropies in the $D^*$ and $B^*$ systems are 5.1\% and 2.6\%
respectively.

Since the $z$-integrated spin asymmetry parameters
in general imply small anisotropies, one might try to get
more sensitivity by
studying the $z$-dependent spin asymmetry parameters defined by:
\begin{eqnarray}
\xi(z,\mu) & = & \frac{D^T(z,\mu)}{D^T(z,\mu) + D^L(z,\mu)} \; , \\
\alpha(z,\mu) & = & \frac{2D^L(z,\mu) - D^T(z,\mu)}{D^T(z,\mu)} \; , \\
{\cal W}(z,\mu) & = & \frac{D^T(z,\mu)}
{D^T(z,\mu) + 2 D^L(z,\mu) }  \; .
\end{eqnarray}
These $z$-dependent spin asymmetry parameters are necessarily
$\mu$ dependent because the fragmentation functions
depend on the scale $\mu$.
They are shown respectively in Fig.~\ref{fig-xiz}(a), (b), and (c) at the
corresponding initial scale $\mu_0$ and $r$ for $J/\psi(r=0.5)$,
$B_c^*(r=0.23)$, $D^*(r=0.17)$, and $B^*(r=0.058)$.
We have chosen $\mu_0$ as
the sum of the heavy constituent quark mass and double the light constituent
quark mass for each bound state, and $r$ to be the ratio of the light
constituent quark mass to the meson mass.
We also take the light $u$ or $d$
constituent quark masses in the $B^*$ and $D^*$ mesons to be 0.3 GeV.
The dependence of $\xi(z)$, $\alpha(z)$, and ${\cal W}(z)$ on $z$ in
Fig.~\ref{fig-xiz}
 shows that the maximum asymmetry occurs at $z$ around 0.6 -- 0.8.
We note that despite having a different initial scale $\mu_0$, the
spin asymmetry parameters for the $\Upsilon$ are the same as those for
the $J/\psi$.

One might worry about how the scale $\mu$ affects the shapes of $\xi(z,\mu)$,
$\alpha(z,\mu)$, and ${\cal W}(z,\mu)$.  Here we remind the readers that the
$z$-integrated $\xi$, $\alpha$, and ${\cal W}$ are
independent of the scale.  We can evolve the polarized
fragmentation functions by solving the
Altarelli-Parisi evolution equation \cite{dglap}
\begin{equation}
\label{evol}
\mu \frac{\partial}{\partial \mu} D^{L,T} (z,\mu) = \int_z^1 \frac{dy}{y} \;
P_{qq}(\frac{z}{y} ) \; D^{L,T}(y,\mu) \; ,
\end{equation}
where $P_{qq}(x) = {\alpha_s (\mu) \over \pi} C_F [(1+x^2)/(1-x)]_+$
is the usual quark-quark Altarelli-Parisi splitting
function, and the plus distribution is defined by
$f(x)_+ = f(x) - \delta(1-x) \int_0^1 dx' f(x')$.
When the initial heavy quark is polarized, we should use the polarized
Altarelli-Parisi splitting function $\Delta P_{qq}(x)$,
which happens to be the same as $P_{qq}(x)$, in Eqn.~(\ref{evol}).
However, in the small $z$ region,
the Altarelli-Parisi evolution equation does not handle the threshold
effect properly \cite{eric}. Unphysical behaviors occur at the small $z$
region for the fragmentation functions and hence for
the spin asymmetry parameters, as one evolves the scale
up to, say, $\mu=m_Z/2$.
Therefore, we can only discuss the evolution behaviors at
the large $z$ region with confidence.
We found that in the large $z$ region the curves for the spin asymmetry
parameters at the scale $\mu = m_Z/2$
do not differ significantly from those given in
Fig.~\ref{fig-xiz}, both in shape and magnitude.
In the future when detailed fragmentation
data for the polarized $D^*$ and $B^*$ become available
at LEP, it will be very interesting to compare our theoretical
predictions of the spin asymmetry parameters
in the large $z$ region given in Fig.~\ref{fig-xiz}
to the experimental results.

A recent set of experimental data on the spin alignment of $D^*$ meson
was from the CLEO detector operating at a center-of-mass energy of 10.5 GeV
\cite{cleo}.  The data set given in Table I of Ref.~\cite{cleo}
is for the spin asymmetry parameter $\alpha(z)$.
The scale of the data set is taken to be half of the
center-of-mass energy, {\it i.e.}, 5.25 GeV, which is not very far from the
$\mu_0=2.1$~GeV of the $D^*$ meson.  We therefore ignore
the evolution effects and directly compare the $D^*$
curve in Fig.~\ref{fig-xiz}(b) with the data from the CLEO measurements.
Unfortunately, since the experimental
data for $\alpha(z)$ had very large error bars, we have to show the
comparison on another graph (Fig.~\ref{fig-cleo}) with a larger vertical
scale.

While our model always predicts a slightly negative value of
$\alpha(z)$ for all values of $z$, the CLEO data
are rather scattered with statistical tendency toward the positive side, as
indicated by the mean value
$\langle\alpha(z)\rangle=0.08\pm 0.07({\rm stat})\pm 0.04({\rm sys})$.
As shown in Fig.~\ref{fig-cleo}, the agreement is good because our curve
$\alpha(z)$ is  within $2\sigma$ of all the data points.
In the  CLEO analysis, they also concluded that the data was marginally
consistent with zero.

In this Section, we have shown that
both the $z$-integrated and the $z$-dependent spin
asymmetry parameters indicate a small misalignment in the
two polarization states of the excited S-wave heavy mesons produced
by heavy quark fragmentation. As indicated by the parameter $\alpha(z)$,
one predicts that the transverse states
produced by heavy quark fragmentation should be populated slightly more than
would be given by naive spin counting over the entire physical region of $z$.

\begin{center}\section{Mean Longitudinal Momentum Fraction}\end{center}

Another useful observable, the mean longitudinal momentum  fraction
$\langle z \rangle$, of the $B_c^*$ (or the heavy-light mesons
$D^*$ and $B^*$, or the heavy quarkonia) at the scale $\mu$ is defined as
\begin{equation}
\label{eq40}
\langle z \rangle = \frac{\int_0^1 dz z D(z,\mu)}{\int_0^1 dz D(z,\mu)} \; .
\end{equation}
In other words, $\langle z \rangle$ is the ratio of the second to the
first moment
of the fragmentation function at the scale $\mu$. Since the
anomalous dimensions of all the moments of the fragmentation function
are known explicitly, the scaling behavior of $\langle z \rangle$ can
be determined as
\begin{equation}
\langle z \rangle = \frac{D(2,\mu)}{D(1)} =
\langle z \rangle_0 \Bigg[ \frac{\alpha_s(\mu)}{\alpha_s(\mu_0)} \Bigg]
^{-{2\gamma \over b}} \; ,
\end{equation}
where $\langle z \rangle_0 = D(2,\mu_0)/D(1)$, $\gamma=-4C_F/3$,
and $b = (11 N_c - 2 n_f)/3$. The first moments $D^{L,T}(1)$
of the longitudinally and transversely polarized fragmentation functions
are given in Eqns.~(\ref{probl}) - (\ref{probt}).
The second moments $D^{L,T}(2,\mu_0)$ at the scale $\mu_0$ are given by
\begin{eqnarray}
\label{mom2l}
D^{L}(2,\mu_0)&=&
\int_0^1 dz z D^L(z,\mu_0) \nonumber \\
&=& 2 N
\Bigg[ {12 + 97r - 453r^2 + 252r^3 + 437r^4 + 15r^5 \over 15\bar r^6}
\nonumber \\
&& \qquad \qquad \qquad \quad
\, + \, {r(5- 11r - 15r^2 + 33r^3 + 12r^4) \over \bar r^7}\log (r)
\Bigg] \; ,
\end{eqnarray}
\begin{eqnarray}
\label{mom2t}
D^{T}(2,\mu_0)&=&
\int_0^1 dz z D^T (z,\mu_0)  \nonumber \\
&=&
4N \Bigg[{12 + 112r -33 r^2 + 252r^3 +17 r^4 \over 15\bar r^6}
\, + \, {r(5 + r + 9r^2+9r^3) \over \bar r^7} \log (r)
\Bigg] \; . \nonumber \\
&& \;
\end{eqnarray}

In Fig.~\ref{zscaling}, we plot $\langle z \rangle^{L,T}$ versus the scale
$\mu$ for  the four meson systems
$J/\psi(r=0.5),\,B_c^*(r=0.23),\,D^*(r=0.17)$, and $B^*(r=0.058)$
with the same input parameters defined in the previous Section.
Due to slightly different value of $\alpha_s(\mu_0)$, the corresponding
curves of $\langle z \rangle^{L,T}$ for the $\Upsilon(r=0.5)$ are slightly
different from those of the  $J/\psi(r=0.5)$. We will not present
the $\Upsilon$ curves here.

As the scale $\mu$ increases across each heavy quark
threshold ($2m_c$ and $2m_b$), the number $n_f$ of active flavors increases by
one unit. The kinks on the curves at $\mu = 2m_c$ and $2m_b$
in Fig.~\ref{zscaling} are due to these threshold effects.
Notice that only slightly noticeable differences between the
longitudinal $\langle z \rangle^L$ and transverse $\langle z \rangle^T$
occur for the $J/\psi$ and $B_c^*$, but
no differences can be seen for the $D^*$ and $B^*$ mesons.
Hence, experimentally using $\langle z \rangle^{L,T}$ to distinguish the
polarizability in heavy quark fragmentation into the S-wave excited
meson states  is not plausible.
We list the $\langle z \rangle_0$ at $\mu=\mu_0$ and
$\langle z \rangle$ at $\mu=m_Z/2$ for the four
different mesons in Table~\ref{table1}.  Numerically, there are no noticeable
difference between the longitudinal and transverse polarizations for all the
four meson systems.
Therefore, to a good approximation, one can set these polarized
 $\langle z \rangle^{L,T}$
values for each meson to be the unpolarized $\langle z \rangle$ value of the
corresponding meson.
Experimentally, unpolarized $\langle z \rangle$ are available from the
LEP, CLEO, and ARGUS data.
The measured quantities are $\langle x_E \rangle_{c\to D^*}$,
$\langle x_E \rangle_{c\to H_c}$, and
$\langle x_E \rangle_{b\to H_b}$,
where $x_E$ is the energy fraction  of the hadron or meson
relative to one half of the center-of-mass energy of the machines, and
$H_c$ and $H_b$ denote the charm and bottom hadrons, respectively. Because
$x_E$ is a good approximation to the fragmentation variable $z$, we simply
treat them to be the same in the following comparisons.
Since the data for $c\to D^*$ is
available separately, we will use $\langle x_E \rangle_{c\to D^*}$
instead of the inclusive $\langle x_E \rangle_{c\to H_c}$. On the other hand,
only the inclusive $\langle x_E \rangle_{b\to H_b}$
has been reported at the LEP.
Nevertheless,
this  inclusive value of $\langle x_E \rangle_{b \to H_b}$ should
be close to the
$\langle x_E \rangle_{b \to B^*}$, since $b \to B^*+X$ is expected to be
the dominant fragmentation mode of the $b$-quark.
In addition, we take the scale of the various
measurements to be one half of the center-of-mass energies of the machines.

Next we will describe briefly how we obtain the average from the LEP, CLEO,
and ARGUS data.
For the LEP data measured values of $\langle x_E \rangle_{c\to D^*}$ are from
OPAL ($0.52 \pm 0.0316$), ALEPH ($0.504 \pm 0.0188$),
and DELPHI ($0.487 \pm 0.0158$) \cite{LEP-c}, in which we
have already combined their systematic and statistical  errors in quadrature
if they are given separately.
We then simply take the mean of the central values from the three experiments
to be the average central value.
For the combined error  we add the absolute errors from the
three experiments in quadrature and divide it by 3.  We thus obtain
$\langle x_E \rangle_{c\to D^*} = 0.504 \pm 0.0133$ by combining
all three LEP data.
Similarly, we have measured values of $\langle x_E \rangle_{b\to H_b}$ from
OPAL ($0.726 \pm 0.023$), ALEPH ($0.67 \pm 0.050$),
DELPHI ($0.695 \pm 0.0326$), and L3 ($0.686\pm 0.017$) \cite{LEP-b}.
Repeating  the same exercise,
we obtain $\langle x_E \rangle_{b\to H_b}=0.694\pm 0.0166$.

The data from CLEO \cite{cleo2} and ARGUS \cite{argus} were given in
form of fragmentation functions.  We need to calculate
$\langle x_E \rangle_{c\to D^*}$ from their
fragmentation data.  In the CLEO paper, the fragmentation function of $c\to
D^*$ were given in Table I (a) and (b) of Ref.~\cite{cleo2},
which correspond to two different detection channels.
We combine the two tables with
the $y$-value of each bin equal to the mean of the two, and the error of the
$y$-value of each bin equal to one half of the
two errors added in quadrature.
Then the value $\langle x_E \rangle$ is obtained by evaluating
the two integrals in the ratio
$\int_{0.37}^1 dx_E x_E D(x_E) / \int_{0.37}^1 dx_E D(x_E)$
by the method of discrete sums.
Assuming the error is only in the $y$-value of each bin, we obtain
$\langle x_E \rangle_{c\to D^*}=0.654 \pm 0.0563$,
where the error is obtained by
adding the errors from each bin in quadrature.
Similarly, from the ARGUS data we obtain
$\langle x_E \rangle_{c\to D^*}=0.642 \pm 0.067$.
Finally, we combine the two averages  from CLEO and ARGUS   and obtain
$\langle x_E \rangle_{c\to D^*}=0.648 \pm 0.0438$.
Since the CLEO and ARGUS operating center-of-mass energies were so close
to each other (10.55 GeV for CLEO and 10.6 GeV for ARGUS), we assume the scale
of both measurements to be 5.3 GeV.

The LEP average for $\langle x_E \rangle_{c\to D^*}$
and $\langle x_E \rangle_{b\to H_b}$ at $\mu=m_Z/2$,
as well as the combined CLEO and ARGUS average for
$\langle x_E \rangle_{c\to D^*}$ at
$\mu=5.3$~GeV are shown in Fig.~\ref{zscaling}.  Excellent agreement between
our predictions and the data is demonstrated.
Here we remind the readers that for the $B^*$ and $D^*$ mesons,
we have assumed a nonrelativistic bound state
picture with the light constituent quark masses being set to be 0.3 GeV.
Uncertainties arising from the overall normalization of the fragmentation
functions are cancelled, as indicated by Eqn.~(\ref{eq40}).
Therefore, the uncertainties in the heavy-light meson systems
come only from how we define the strong coupling constant $\alpha_s$,
the initial scale $\mu_0$ for the fragmentation functions,
and what values we choose for the light constituent quark masses.
Summing up,  the mean longitudinal
momentum fractions $\langle z \rangle^{L,T}$ for a heavy quark to fragment
into polarized vector mesons do not show any measurable difference
between the longitudinal and transverse components.
Nevertheless, the predictions of $\langle z \rangle$ by our
fragmentation functions
for $D^*$ and $B^*$ mesons at different scales are in excellent
agreement with the measured data from  the LEP, CLEO, and ARGUS detectors.

\begin{center}\section{Transverse Momentum $p_\perp$ Dependence}\end{center}

Sections III and IV showed that
the ratio of longitudinal-to-transverse populations from
$\bar b$-quark  fragmentation into polarized $B_c^*$ mesons
is only marginally different from  that given by the naive spin counting
({\it i.e.} the heavy quark mass limit), as indicated
by both the $z$-integrated and $z$-dependent spin asymmetry parameters (see
Figs.~\ref{fig-xi} and \ref{fig-xiz}).  Similar conclusions hold for the other
mesons $J/\psi$, $\Upsilon$, $D^*$, and $B^*$.
As a consequence, the two body decays of
the excited S-wave mesons into the corresponding pseudoscalar ground states
with emission of photon or pion are
almost isotropic.
In addition, we showed in Section V that it is not feasible to use the
mean longitudinal momentum fractions $\langle z \rangle^{L,T}$
to distinguish the longitudinal and  transverse
polarizations of the excited S-wave mesons
produced by fragmentation of heavy quarks (see Fig.~\ref{zscaling}).
Hence, information about the spin of a heavy quark is not easy to
extract from the fragmentation data of the heavy quark into
S-wave excited mesons \cite{note}.

So far, we have only investigated the $z$-dependence of the fragmentation
functions and of the spin asymmetry parameters.  All
the dependence on the motion of the meson perpendicular to the
fragmentation axis has been integrated out.
In this Section we will investigate the dependence of the fragmentation
functions on the $p_\perp=|\vec p_\perp|$, where $\vec
p_\perp =(p_1, p_2)$ is the transverse momentum of the meson with respect
to the fragmentation axis.
Recall that from Eqn.~(\ref{perp}), we have
\begin{equation}
p^2_{\perp} = z(1-z) \left(
s - \frac{M^2}{z} - \frac{m_c^2}{1-z} \right) \; .
\label{ptands}
\end{equation}
Introducing the dimensionless variable $t = p_\perp /M$ and trading
the variable $s$ to $t$ from Eqn.(\ref{ptands}), we can define the
fragmentation functions $D(z,t)$ and $D(t)$ according to
\begin{eqnarray}
\label{dztdt}
\int_0^{\infty} dt D(t) &=&
\int_0^1 dz \int_0^{\infty} dt D(z,t) \;\; , \nonumber \\
& = & \int_0^1 dz \int ds
\; \theta \left( s - {M^2 \over z} - {m_c^2 \over 1-z} \right) \;
D(z,s) \;\; .
\end{eqnarray}
This implies
\begin{equation}
\label{dzpt}
D(z,t) = {2M^2t \over z(1-z)} D(z,s) \; \; , \; \; {\rm with} \; \;
s=M^2 \Bigg[ \frac{1+t^2}{z}+\frac{r^2+t^2}{1-z} \Bigg] \; .
\end{equation}
To our knowledge, QCD evolution equation for
the fragmentation function $D(z,t)$
that depends on both the longitudinal momentum fraction $z$ and the rescaled
transverse momentum $t$ has not been written down. But formalism, like those
in Refs.~\cite{ktfac}, which dealt with similar issue in the parton
distribution functions, may apply for fragmentation functions as well.
For the following we will ignore the issue of QCD evolution effects
in the fragmentation functions with explicit $t$-dependence.

Integrating over $z$ in Eqn.~(\ref{dzpt}),  we obtain the $t$-dependent
polarized fragmentation functions $D^{L,T}(t)$:
\begin{eqnarray}
\label{dtl}
D^L(t,\mu_0) &=& \frac{Nr}{2\bar r^6}\frac{1}{t^6(1+t^2)^2}
\Bigg\{ 24 \bar r^2 t^7 (1+t^2)^2 \log(r)    \nonumber \\
& + & 12 t (1+t^2)^2 \Bigg[ 4r^3 - r(2+r+2r^2)t^2 + \bar r^2 t^6 \Bigg]
\log \left( {1+ {t^2 \over r^2} \over 1+t^2} \right)  \nonumber \\
& + & 3 (1+t^2)^2  \Bigg[ 10r^4-r^2(33+20r-8r^2)t^2
	+(3+2r-5r^2+12r^3+8r^4)t^4  \nonumber \\
&& \qquad \qquad   +(5-14r+4r^2+8r^3) t^6 \Bigg]
	{\rm arctan}\left( {\bar r t \over r+t^2} \right) \nonumber \\
& - & \bar r t
	\Bigg[ 30r^3 -r(61+22r-74r^2)t^2+(5-146r-38r^2+80r^3)t^4 \nonumber \\
&& \qquad \quad   \left.
+(4-103r-4r^2+52r^3)t^6 + 3(1-10r+8r^2+4r^3)t^8 \Bigg] \right\} \, ,
\end{eqnarray}
\begin{eqnarray}
\label{dtt}
D^T(t,\mu_0) &=& \frac{Nr}{\bar r^6}\frac{1}{t^6(1+t^2)^2}
\left\{ 12r t (1+t^2)^2 \Bigg[ 4r^2 - (2+3r)t^2 \Bigg]
\log \left( {1+ {t^2 \over r^2} \over 1+t^2} \right) \right. \nonumber \\
& + & 3(1+t^2)^2  \Bigg[ 10r^4-r^2(33+23r-11r^2)t^2
	+(3+9r+3r^2+5r^3)t^4  \nonumber \\
&& \qquad \qquad \qquad  +(2+r) t^6 \Bigg]
	{\rm arctan}\left( {\bar r t \over r+t^2} \right) \nonumber \\
& - & \bar r t
	\Bigg[ 30r^3 -r(61+31r-83r^2)t^2+(5-122r-53r^2+71r^3)t^4 \nonumber \\
&& \qquad \qquad \qquad \left.
+(13-64r-16r^2+16r^3)t^6+3(2+r)t^8 \Bigg] \right\} \; .
\end{eqnarray}
As a cross check, one can integrate $D^{L,T}(t,\mu_0)$ over $t$ from
0 to $\infty$,
and get back the total fragmentation probabilities given in
Eqns.~(\ref{probl}) and (\ref{probt}).

A couple of asymptotic behaviors of
$D^{L,T}(t,\mu_0)$ are in order.
As $t \to 0$, $D^{L,T}(t,\mu_0)$ vanishes linearly in $t$:
\begin{eqnarray}
D^L(t,\mu_0) & \to  & {2Nrt \over \bar r^4}
\Bigg[ 6 \log(r) + {\bar r \over 35r^3}
(8-46r+160r^2+101r^3-13r^4) \Bigg] \; , \\
D^T(t,\mu_0) & \to & {8(4+3r)Nt \over 35 r^2} \; .
\end{eqnarray}
As $t \to \infty$, $D^{L,T}(t,\mu_0)$ falls off like $1/t^3$
according to
\begin{eqnarray}
D^L(t,\mu_0) & \to & {Nr \over \bar r^6 t^3}
\Bigg[ 12 r(2+r+2r^2)\log(r) + \bar r (1+r) (6+20r+r^2+4r^3-r^4)
\Bigg] \; , \nonumber \\
&& \\
D^T(t,\mu_0) & \to & {2 Nr \over \bar r^6 t^3}
\Bigg[ 12 r(2+3r)\log(r) + \bar r (3 + 47r + 11r^2 - r^3) \Bigg]  \; .
\end{eqnarray}

The dependence of the fragmentation functions $D^{L,T}_{\bar b\to
B_c^*}(t,\mu_0)$ on $t$ at the initial scale $\mu_0$
are  shown in Fig.~\ref{dperp}.
Both functions peak at $t \approx 0.2$, {\it i.e.},
$p_\perp \approx 1.3$~GeV for $B_c^*$ meson.
At large $p_\perp$, one sees that
the transverse $D^T(t)$ falls off more rapidly than the longitudinal $D^L(t)$.
Explicitly, the
curves for the longitudinal and transverse  polarizations cross over at
$t \approx 1.5$, {\it i.e.}, $p_\perp \approx 10$~GeV for $B_c^*$ mesons.
This is a very clean sign of
difference between the longitudinally and transversely polarized states of
$B_c^*$.
Overall, the transverse states are populated about twice as much
as the longitudinal one, but the longitudinal component becomes dominant
beyond the cross-over.
Unfortunately, it might be hard to observe the cross-over experimentally
because the cross-over lies well beyond their peak values.
Both the longitudinal and transverse spectra have dropped  substantially
by the time they reach the cross-over,
which implies very small cross sections for $p_\perp > 10$~GeV.
In addition, there are large uncertainties in determining the fragmentation
axis of a jet and consequently also the value of $p_\perp$ for the $B_c^*$.
Nevertheless, in principle, imposing a high $p_\perp$ cut
can eliminate a large sample of the transverse population of the $B_c^*$.

We can also define the average transverse momentum $\langle p_\perp \rangle$ by
\begin{equation}
{\langle p_\perp \rangle \over M} = \langle t \rangle =
\frac{\int_0^\infty dt \, t \, D(t)}{\int_0^\infty dt \,
D(t)} \; .
\end{equation}
Numerically, the average $\langle t\rangle$ for the longitudinal and transverse
$B_c^*$ are 1.1 and 0.61 respectively, and so  the corresponding average
$\langle p_\perp \rangle$ are 7.0 GeV and 3.9 GeV, respectively.
Unlike $\langle z \rangle$, there is a big difference in
$\langle p_\perp \rangle$ between the longitudinal and transverse
states of $B_c^*$.
The cross-over, and the average $\langle t\rangle$ and
$\langle p_\perp \rangle$ for
the $B_c^*$, $J/\psi,\,D^*$, and $B^*$  are all summarized in Table
\ref{table2}.
Entries for $\Upsilon$ are the same as those for $J/\psi$, except that
the average $p_\perp$ for $\Upsilon$ is larger by a factor of $m_b/m_c$.
A significant difference in $\langle t\rangle$ between the longitudinal and
transverse states persist in $J/\psi,\,\Upsilon,\,D^*$, and $B^*$ systems.

Despite the difficulties in their experimental measurements,
it is instructive to evaluate the $t$-dependent
spin asymmetry parameters defined by
\begin{eqnarray}
\label{xit}
\xi(t) &=& \frac{D^T(t)}{D^T(t)+D^L(t)} \; , \\
\label{alphat}
\alpha (t) &=& \frac{2D^L(t)-D^T(t)}{D^T(t)} \; , \\
\label{wt}
{\cal W} (t) &=& \frac{D^T(t)}{D^T(t) + 2 D^L(t)} \; .
\end{eqnarray}
In Fig.~\ref{fig-xiperp}, we plot
the three spin asymmetry parameters $\xi(t),\,\alpha(t)$, and ${\cal W}(t)$
as functions of $t$ for the mesons that we are considering.
The curves in Fig.~\ref{fig-xiperp} vary far more
dramatically  than the corresponding curves in Fig.~\ref{fig-xiz}, because the
longitudinal and transverse fragmentation functions show a very different
dependence on $p_\perp$. For all the vector meson systems that
we are considering,
the $t$-dependent longitudinal and transverse fragmentation functions
cross over at $t \approx 1.5-2.0$.
Consequently, the corresponding spin asymmetry parameters
also change very rapidly at $t\approx 1.5-2.0$.

With $D^{L,T}(t)$ replaced by $D^{L,T}(z,t)$ in
Eqns.(\ref{xit}) - (\ref{wt}),
one can also introduce the spin asymmetry parameters
$\xi(z,p_\perp)$, $\alpha (z,p_\perp)$, and ${\cal W}(z,p_\perp)$
that depend on both the variables $z$ and $p_\perp$.
It may be possible to measure such spin asymmetry parameters,
we will not discuss them any further here.

\begin{center}\section{Conclusions}\end{center}

In this paper, we have used PQCD to derive the polarized fragmentation
functions  to leading order in $\alpha_s$ for a $\bar b$ antiquark fragmenting
into the longitudinally and  transversely polarized $B_c^*$ states.
Parity invariance also implies  that the polarized heavy quark fragmentation
functions $D^L(z)$ and $D^T(z)$ derived in this paper are the same whether
the initial heavy quark is polarized or unpolarized.
These polarized fragmentation functions can be used to define various spin
asymmetry parameters that can be determined  experimentally.

In this work, we have also used the PQCD fragmentation functions to study
the spin alignment of the other S-wave excited mesons that carry charm or
beauty.
The spin asymmetry parameter $\alpha(z)$ for $c \to D^*$ is consistent
with the measurements by the CLEO detector.
The mean longitudinal momentum fractions $\langle z \rangle$
predicted for  $D^*$ and $B^*$ are also in excellent
agreement with the measurements of the LEP, CLEO, and ARGUS detectors.
We also demonstrate a very interesting dependence of the polarized
heavy quark fragmentation functions and the spin asymmetry parameters on the
transverse momentum $p_\perp$ of the vector meson relative to the
fragmentation axis.  Longitudinally polarized vector mesons have a harder
$p_\perp$ spectrum  than the transversely polarized states.

The perturbative QCD fragmentation functions that we obtained in this paper
and in Refs.~\cite{zpaper,bcfrags} are expected to work well for
$c \to \eta_c, J/\psi$, $b \to \eta_b,\Upsilon$, and $\bar b \to B_c, B_c^*$ in
any high energy processes  with large transverse momentum $p_T$.
They are rigorously correct in the limit that the heavy quark masses
$m_c$ and $m_b$ are much larger than $\Lambda_{\rm QCD}$.    Corrections to
the fragmentation contributions are of order $M^2/p_T^2$ ($M$ is the mass of
the meson) and therefore small at large enough $p_T$.

Our fragmentation functions also seem to provide a successful phenomenological
model  for describing the spin dependence of charm and bottom fragmentation
into heavy-light mesons like $D^*$ and $B^*$. Furthermore,  our fragmentation
functions  are  consistent with  heavy quark symmetry.
Dominant error comes from the neglect of relativistic corrections and higher
order perturbative corrections.
We conclude  that the PQCD-inspired fragmentation functions could be useful
in describing charm and bottom fragmentation in the future experiments.

\bigskip

\section*{Acknowledgements}

We are grateful to Eric Braaten for useful discussions and careful reading of
the manuscript, and to Peter Heimberg for a discussion on calculating the
experimental errors presented in Section V.
This work was supported by the U.~S. Department of Energy, Division of
High Energy Physics, under Grants DE-FG02-91-ER40684 and DE-FG03-91ER40674 and
by Texas National Research Laboratory Grant RGFY93-330.
%-----------------------------------

\vfill\eject

%-------------------------------------
%\begin{references}

%%%%%%%%%%%%%%%
\begin{table}[p]
\centering
\caption[]{\label{table1}
\small
The mean longitudinal momentum fraction $\langle z \rangle$ when a heavy quark
fragments into a polarized heavy meson in the spin-orbital state $^3S_1$
at the scales $\mu=\mu_0$ and $\mu=m_Z/2$.  We take $m_b=4.9$ GeV, $m_c=1.5$
GeV, and the light $u$ or $d$ constituent quark masses to be 0.3 GeV.
Here L stands for longitudinal and T for transverse.
}
\medskip
\begin{tabular}{|c@{\extracolsep{0.3in}}cc|cc|cc|}
\hline
\hline
Meson   &   $r$  & $\mu_0$  & \multicolumn{2}{c|}
{$\langle z \rangle_{\mu=\mu_0}$}  &
\multicolumn{2}{c|}{$\langle z \rangle_{\mu=m_Z/2}$}  \\
\hline
&&& \underline{L} & \underline{T} & \underline{L} & \underline{T} \\
$J/\psi$   &  0.50  & 4.5  &  0.61 & 0.62 & 0.48 & 0.48 \\
$B_c^*$    &  0.23  & 7.9  &  0.73 & 0.73 & 0.61 & 0.61 \\
$D^*$      &  0.17  & 2.1  &  0.77 & 0.77 & 0.50 & 0.50 \\
$B^*$      &  0.058 & 5.5  &  0.87 & 0.87 & 0.70 & 0.70 \\
\hline
\end{tabular}
\end{table}

%%%%%%%%%%%%%%%%%%%%%%%%%%%%%%%%%%%%%%%
\newpage
\begin{table}[p]
\centering
\caption[]{\label{table2}
\small
A table showing where the curves $D^L(t)$ and $D^T(t)$ cross-over, the average
$\langle t\rangle=\langle p_\perp \rangle/M$,
and the average $\langle p_\perp \rangle$ for the $\bar b$ quark
fragmenting into $B^*$ and $B_c^*$, and for the $c$ quark
fragmenting into $D^*$ and $J/\psi$.
The mass of each meson is taken to be the sum of the
constituent quark masses ($m_b=4.9,\,m_c=1.5,\,m_{u,d}=0.3$ GeV).
}
\medskip
\begin{tabular}{|c@{\extracolsep{0.3in}}cccccc|}
\hline
\hline
         & $r$ & $t$(cross-over)  & \multicolumn{2}{c}{$\langle t\rangle$} &
\multicolumn{2}{c|}{$\langle p_\perp \rangle$ (GeV)} \\
\hline
         &      & & \underline{L}   & \underline{T} &  \underline{L}  &
\underline{T} \\
$J/\psi$ & 0.5  & 1.6          & 1.8 & 0.86         & 5.5 & 2.6 \\
$B_c^*$ & 0.23  & 1.5         & 1.1  & 0.61         & 7.0 & 3.9 \\
$D^*$ & 0.17 & 1.6           &  0.86 & 0.52         & 1.5 & 0.94 \\
$B^*$ & 0.058 & 2.0          &  0.41 & 0.30         & 2.1 & 1.5  \\
\hline
\end{tabular}
\end{table}

%%%%%%%%%%%%%%%%%%%%%%%%%%%%%%%%%%%%%%%%%%%%%%%%%%%%%%%%%%%%%%%%%
\newpage
\begin{center}\section*{Figure Captions}\end{center}

\begin{enumerate}

\item
\label{fig-feynman}
The leading order Feynman diagram contributing to the fragmentation process
$\bar b \to B_c^*$.

\item
\label{fig-Dz}
The polarized fragmentation functions $D^L_{\bar b\to B_c^*}(z)$ and
$D^T_{\bar b\to B_c^*}(z)$ versus $z$ at the initial scale $\mu_0$.
The sum of the first two terms in the heavy quark mass expansion
are also shown for the longitudinal and transverse polarizations.

\item
\label{fig-xi}
The spin asymmetry parameters
(a) $\xi$, (b) $\alpha$, and  (c) ${\cal W}$ versus the mass ratio
parameter $r$.

\item
\label{fig-xiz}
The spin asymmetry parameters
(a) $\xi(z,\mu)$, (b) $\alpha(z,\mu)$, and (c) ${\cal W}(z,\mu)$
versus $z$ at the initial scale $\mu_0$ for
$r=0.5$ (solid), 0.23(dash), 0.17 (dotdash), and 0.058 (dot)
corresponding to $J/\psi$, $B_c^*$, $D^*$, and $B^*$, respectively.

\item
\label{fig-cleo}
A comparison between the spin asymmetry parameter $\alpha(z)$ for $c\to D^*$
predicted by our model and the experimental measurements
by the CLEO Collaboration \cite{cleo}.

\item
\label{zscaling}
The mean longitudinal momentum fraction
$\langle z \rangle^{L,T}$ versus the scale $\mu$
at $r=0.5, 0.23, 0.17$, and $0.058$ for the $J/\psi$,  $B_c^*$, $D^*$,
and $B^*$, respectively.
The longitudinal curve is solid and the transverse one is dotted.
The two measurements $\langle x_E \rangle_{c\to D^*}$ and
$\langle x_E \rangle_{b\to H_b}$ from the LEP detectors are shown at
$\mu=m_Z/2$, and the combined CLEO and ARGUS measurement
$\langle x_E \rangle_{c\to D^*}$ is at $\mu=5.3$~GeV.

\item
\label{dperp}
$D^L_{\bar b \to B_c^*}(t)$ and $D^T_{\bar b \to B_c^*}(t)$
versus the rescaled transverse momentum $t=p_\perp/M$ of the $B_c^*$ meson
at the initial scale $\mu_0$.

\item
\label{fig-xiperp}
The spin asymmetry parameters
(a) $\xi(t)$, (b) $\alpha(t)$, and (c) ${\cal W}(t)$
versus $t=p_\perp/M$ at $r$=0.5 (solid), 0.23 (dash), 0.17 (dotdash),
and 0.058 (dot), which correspond to
$J/\psi$, $B_c^*$, $D^*$, and $B^*$ mesons, respectively.

\end{enumerate}


\begin{thebibliography}{99}
\bibitem{note}
{Falk and Peskin in Ref.\cite{falkpeskin} have
pointed out that in the heavy quark limit
heavy quark fragmentation into P-wave excited states can in general
lead to anisotropies in the decay products of these excited states.}
\end{thebibliography}
\end{document}